\documentclass[prl,twocolumn,showpacs,superscriptaddress,oneside]{revtex4}%
\usepackage{amssymb}
\usepackage{amsfonts}
\usepackage{graphicx}
\usepackage{epsfig}
\usepackage{amsmath}
\begin{document}
\title{Hall effect and resistivity study of the magnetic transition, carrier content
and Fermi liquid behavior in Ba(Fe$_{1-x}$Co$_{x}$)$_{2}$As$_{2}$}
\author{F. Rullier-Albenque}
\email{florence.albenque-rullier@cea.fr}
\affiliation{Service de Physique de l'Etat Condens\'e, Orme des Merisiers, CEA Saclay (CNRS URA 2464), 91191 Gif sur Yvette cedex, France}
\author{D. Colson}
\affiliation{Service de Physique de l'Etat Condens\'e, Orme des Merisiers, CEA Saclay (CNRS URA 2464), 91191 Gif sur Yvette cedex, France}
\author{A. Forget}
\affiliation{Service de Physique de l'Etat Condens\'e, Orme des Merisiers, CEA Saclay (CNRS URA 2464), 91191 Gif sur Yvette cedex, France}
\author{H. Alloul}
\affiliation{Laboratoire de Physique des Solides, UMR CNRS 8502, Universit\'e Paris Sud, 91405 Orsay, France}

\date{30 March 2009, published 30 July 2009}

\begin{abstract}
The negative Hall constant $R_{H}$ measured all over the phase diagram of
Ba(Fe$_{1-x}$Co$_{x}$)$_{2}$As$_{2}$ allows us to show that  electron
carriers always dominate the transport properties. The evolution of $R_{H}%
$ with $x$ at low doping ($x<2\%$) indicates that important band structure
changes happen for $x<2\%$ prior to the emergence of superconductivity. For
higher $x$, a change with $T$ of the electron concentration is required to explain the low $T$
variations of $R_{H}$, while the electron scattering rate displays the
$T^{2}$ law expected for a Fermi liquid. The $T=0$ residual scattering is
affected by Co disorder in the magnetic phase, but is rather dominated by
incipient disorder in the paramagnetic state.
\end{abstract}

\maketitle

\paragraph{Introduction.}
Despite the similarity of phase diagrams of cuprates and iron based pnictides
\cite{Zhao, Chen, Chu}, it is already well established that the underlying
physics of these compounds is different. While cuprates are doped Mott
insulators with properties tightly related to the effects of strong electron
correlations in the narrow Cu $3d$ band, iron pnictides are metallic systems
with spin density wave (SDW) state at low doping. They display a semi-metal
band structure with three hole bands and two electron bands close to the Fermi
level $E_{F}$ as obtained by electronic structure calculations \cite{Singh,
Mazin1, Ma} and confirmed by angular resolved photoemission spectroscopy
(ARPES) \cite{Hsieh, Yi}. The SDW ordering is often attributed to the
nesting between electron and hole cylindrical bands. The interband interaction
between the quasi nested bands could be as well at the origin of the
high-$T_{c}$ values \cite{Mazin2}. The proximity and/or coexistence
of antiferromagnetism and superconductivity (SC) in the Fe-pnictides 
and the detection of spin fluctuations by NMR \cite{Ning} have been
also taken as an indication that magnetic fluctuations may play a decisive role
in the SC pairing mechanism as proposed for the high-$T_{c}$ cuprates. 

Surprisingly, although some linear dependences of resistivity have been taken as indications for a quantum critical point \cite{Gooch}, systematic analyses of the transport properties to reveal
features expected for such a band structure have not been
performed so far all over the phase diagram. Presently such studies appear accessible for the Ba(Fe$_{1-x}$Co$_{x}$)$_{2}$As$_{2}$ family as fine tuning of
Co content can be achieved in sizeable single crystals, allowing then to span
the entire electron doped phase diagram \cite{Sefat, Leithe, Chu, Ni, X.F.Wang2}. 
Investigations of the influence of Co disorder in Fe
layers are still missing, even if it has been pointed out \cite{Leithe} that
Co substitution has less incidence on SC than in plane substitutions in the
cuprates \cite{RMP}. We address these points in this work by performing
extensive measurements of resistivity $\rho$ and Hall
coefficient $R_{H}$, similar to those reported simultaneously in \cite{Fang}. Here the great accuracy 
of our data allows us to demonstrate that the hole contribution to the
transport can be neglected at low $T$ in most of the phase diagram. 
This leads us to propose that in this pnictide family the electron carriers have
archetypal Fermi liquid behavior, with large $T^{2}$ contributions
to the electron scattering rate, as $k_{B}T$ can become sizeable with respect
to $E_{F}$. The hole contribution to the transport is found anomalous and only perceptible at high $T$ for low $x$. At the magnetic transition $T_{SDW}$ of the undoped parent compound, $R_{H}$ is found to evolve in two steps which appear to correspond to the successive gapping of the electronic bands with decreasing $T$.

\paragraph{Samples and resistivity measurements.}
Single crystals of Ba(Fe$_{1-x}$Co$_{x}$)$_{2}$As$_{2}$ with Co contents $x$
ranging from 0 to 0.2 were grown using the self-flux method \cite{X.F.Wang1}. Starting reagents of high purity Ba, FeAs and CoAs were
mixed in the molar ratio $1$:$(4-x)$:$x$, loaded in alumina crucibles and then
sealed in evacuated quartz tubes. The mixture was typically held at $1180$\ensuremath{^\circ}C for 4 hours, slowly cooled down to $1000$\ensuremath{^\circ}C/h at 5\ensuremath{^\circ}C/h, then down to room temperature at 200\ensuremath{^\circ}C/h. Single crystals could be extracted mechanically from the solid flux.
Chemical analyses were performed with an electron probe (Camebax 50), on several crystals for each Co doping, yielding the Co content within 0.5\% absolute accuracy. In-plane resistivity measurements using either the standard dc
four-probe or the van der Pauw technique \cite{VdP} were performed on samples
which have been cleaved from larger crystals to thicknesses lower than $30$
$\mu m$. Two or three different samples were measured for each Co doping and a
good reproducibility of the data was found both for the $T$ dependence of
$\rho$ and its absolute values (better than 10\% accuracy, whatever the technique used).

\begin{figure}
\centering
\includegraphics[width=8cm]{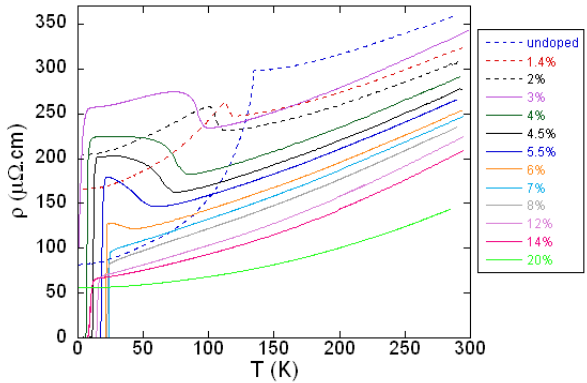}
\caption{(color on line) Temperature dependence of
the in-plane resistivity $\rho(T)$ of Ba(Fe$_{1-x}$Co$_{x}$)$_{2}$%
As$_{2}$ single crystals. For the sake of clarity, the data for the undoped
parent and low doped non SC samples, which evolve differently than for higher Co content, 
are plotted with dashed lines.
Notice the singularities associated with the SDW transitions.}
\label{Fig.1}
\end{figure}

The qualitative features of our $\rho(T)$ data displayed in fig.1 on a series
of samples match reported results \cite{Leithe, Chu, Ni}: the drop of
resistivity at 135K, which signals the structural and SDW transitions in the
undoped parent, is replaced by a step like increase of $\rho(T)$ as soon as Co
is inserted. Upon Co doping this transition is shifted to lower $T$ and SC is
observed within the doping range $0.03\lesssim x\lesssim0.15$, showing the
coexistence of SC and SDW order for $ 0.03\lesssim x\lesssim0.07$.
We notice that a marked change in the evolution of $\rho(T)$ curves occurs at about $x=0.03$ 
for which SC appears. In both the paramagnetic and SDW phases $\rho$ only starts there to decrease monotonously with Co content. Another puzzling observation concerns the $T$
dependences of $\rho$ in the paramagnetic phase which parallel each other for most Co dopings. This seems difficult to conciliate with a multiband description of the
electronic structure for which the conductivity $\sigma$ is the sum of the hole and electron contributions ($\sigma=1/\rho=\sigma_{e}+\sigma_{h}$). How could
the evolutions with $x$ of the concentrations and relative
mobilities of the two types of carriers compensate and give similar $T$ dependent contributions to $\rho$ whatever $x$?

\begin{figure}
\centering
\includegraphics[width=8cm]{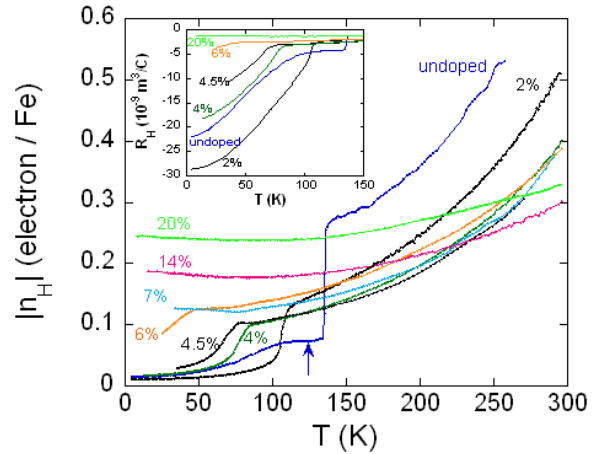}
\caption{(color on line) $T$ variation of the Hall
number $|n_{H}|$ for a set of samples. Here, $n_{H}$ (\textit{e}/Fe)
$=0.32/R_{H}x(10^{-9}m^{3}/$C). For the undoped compound, the drop below $T_{SDW}$ occurs in two steps, the value on the plateau indicated by an arrow being 0.074 \textit{e}/Fe. 
Inset: Raw data for the magnetic samples showing the drops of
$R_{H}$ at $T_{SDW}$. The data for the most overdoped sample have
been kept for comparison.}
\label{Fig.2}
\end{figure}

\paragraph{Hall effect.}
In order to get more insight into the evolution of transport properties, we
have then performed Hall effect measurements on the same samples, and checked the linearity in $H$ of the Hall voltage up 8T in the paramagnetic phase. The data for the Hall coefficient $R_{H}$ are presented in the inset of fig.2 for a set of Co dopings. The strong increases in the magnitude of $R_{H}$ at the SDW transitions will be discussed later. In order to better visualize
the behavior of $R_{H}$ in the paramagnetic phases and for large $x$, we
have plotted in fig.2 the variations of the Hall number
$n_{H}=1/(eR_{H})$. Assuming a simple two band model, one could write:
\begin{equation}
\label{Eq.1}
eR_{H}=\frac{1}{n_{H}}=\dfrac{\sigma_{h}^{2}}{n_{h}(\sigma_{e}+\sigma
_{h})^{2}}-\dfrac{\sigma_{e}^{2}}{n_{e}(\sigma_{e}+\sigma_{h})^{2}}
\end{equation}
where $n_{e}$ ($n_{h}$) are the concentration of electrons (holes) usually taken as $T$ independent in a metallic state. One should note that charge conservation implies:
\begin{equation}
\label{Eq.2}
n_{e}=n_{h}+x, 
\end{equation}
with the usual assumption that Co gives one electron to these bands. As we
have only three relations for $n_{e},n_{h},\sigma_{e},\sigma_{h}$, the problem
cannot be readily solved without further elements based on physical
arguments or other experiments. However the negative sign of $R_{H}$
indicates that electrons give the dominant contribution to the charge
transport at all $T$ whatever $x$ with the value of $|n_{H}|$ being an upper limit for $n_{e}$.

For large Co doping, ARPES
experiments show that the hole Fermi surface pockets become very small
while the electron pockets significantly expand \cite{Sekiba}. In this
case $n_{h}\ll n_{e}$ (and $\sigma_{h}\ll\sigma_{e}$) so
that Eq.\ref{Eq.1} writes:
\begin{equation}
\label{Eq.3} 
|n_{H}|\simeq n_{e}\left( 1+2\sigma_{h}/\sigma\right)
\end{equation}
to first order in $\sigma_{h}/\sigma$. This simple limit permits us to evidence that, whatever the exact value of $n_{e}$, the data for $\rho(T)$ and $|n_{H}(T)|$ would only be explained by a high $T$ increase of $\sigma_{h}$ with a weak minimum below 100K. This behavior, opposite to that
expected for a metallic system, leads us to conclude that $n_{e}$
and $n_{h}$ are $T$ dependent. 
\begin{figure}
\centering
\includegraphics[width=8cm]{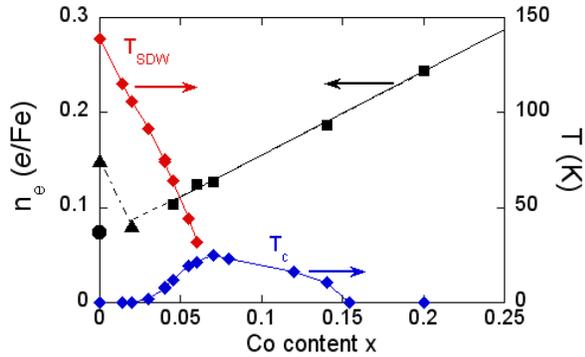}
\caption{(color on line) Phase diagram of Ba(Fe$_{1-x}$Co$_{x}$)$_{2}$As$_{2}$. 
$T_{SDW}$ has been taken at the minimum of
$d\rho/dT$ \cite{Pratt}. ($\blacksquare$) Values of $n_{e}$ determined at low $T$ for $x\geq4$\%. The linear fit (full line) extrapolates for $x=0$ to 0.07, close to the value ($\bullet$) of the intermediate plateau of $|n_{H}(T)|$ (arrow in fig.2). Extrapolations from the paramagnetic phase ($\blacktriangle$), assuming $T^{2}$ variations for $n_{H}(T)$ as found for higher $x$, allow
us to conjecture the evolution of $n_{e}$ (dash line) which would take place below $x=0.03$ in the absence of magnetic ordering.}
\label{Fig.3}
\end{figure} 

For the most doped samples for which hole contributions to both
$n_{H}(T)$ and $\rho$ $(T)$ can be neglected, these quantities
resume then into the single band expressions:
\begin{equation}
 \label{Eq.4}
|n_{H}(T)|=n_{e}(T)\ \text{and }\rho(T)=\frac{m_{e}}{n_{e}(T)e^{2}%
\tau(T)} ,
\end{equation}
where $m_{e}$ is the electron effective mass. The limiting values of
$n_{e}$ at $T=0$ deduced from the low $T$ data for $|n_{H}|$ are plotted in
fig.3 versus Co content. In this plot we have reported as well the data
for the magnetic samples for which $|n_{H}|$ flattens out just above
$T_{SDW}$. We find that
$|n_{H}(T=0)|$ increases linearly with $x$ which supports the idea that the low $T$ value of $|n_{H}|$ can be assimilated to the actual value of $n_{e}$ down to $x\simeq4$\%. Each Co adds $\sim0.9$ \textit{e}/Fe and only reduces $n_{h}$ from 0.06 to 0.04 hole/Fe for $0.04\leq x\leq0.2$. In the simple rigid band model this trend can be expected, as the shift of $E_{F}$ yields different changes of $n_{e}$ and $n_{h}$ if shapes of the pockets (and/or effective masses) are different for the two types of carriers, as evidenced by ARPES \cite{Yi, Sekiba}. 

As for the small $T$ variation of the carrier content for $T\leq150$K, it can be assigned to the fact that 
$E_{F}$ is only 20-40meV above the bottom of the electron bands in the BaFe$_{2}$As$_{2}$ family as shown by ARPES \cite{Yi, Yang}. Thermal population of the hole and electron
bands yields then a shift of the chemical potential $\mu (T)$ to
fulfill Eq.\ref{Eq.2}. However this band effect might not be sufficient to explain the huge increase of $n_{H}$ by a factor 3 to 5 observed above 150K for $x\leq7$\%. Here we might consider that the smallness of $\sigma_{h}$ at low $T$ could result from the localization of holes in low energy states (such as defects). The $T$ variations of $\mu$ and of the carrier contents would then result from the high $T$ release of such states.
\begin{figure}
\centering
\includegraphics[width=8cm]{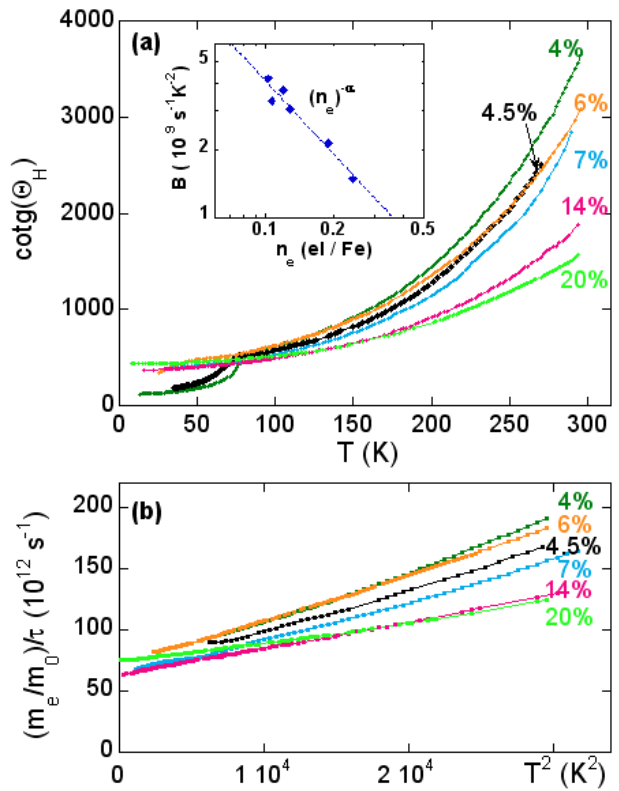}
\caption{(color on line)(a) $T$ variation of cotg$(\Theta_{H})=\rho/|R_{H}|$ at 1T. (b) This quantity, which reduces to $(m_{e}/m_{0}/)/\tau$ if $\sigma_{h}\ll\sigma_{e}$, is plotted versus $T^{2}$ in the paramagnetic phase for $x\gtrsim4$\%. This evidences a Fermi liquid behavior for $T\leq150K$. 
Inset: Coefficient $B$ of the $T^{2}$ variation versus $n_{e}$ in logarithmic scales. The line corresponds to $B\propto1/n_{e}$.} 
\label{Fig.4}
\end{figure}

\paragraph{Electron scattering rate.}
So far we have interpreted satisfactorily the low $T$ Hall data assuming that the holes do not
contribute to the charge transport. The quantity cotg$(\Theta_{H})=\rho/
|R_{H}|$ equals $m_{e}/e\tau$ in the single band case of Eqs.\ref{Eq.4}, and gives a direct determination of the electron scattering rate $1/\tau(T)$ when this limit applies. We have then reported in Fig.4(a) the data for cotg$(\Theta_{H})$ for all our superconducting samples. The very similar behavior displayed by cotg$(\Theta_{H})$ for all $x$ does not show up any \textit{direct} incidence on the transport of the spin fluctuations detected above $T_{SDW}$ in the nuclear spin lattice relaxation rates \cite{Ning}. This might lead us to anticipate that the variation of $n_{H}$ is always governed by a $T$ variation of $n_{e}$, even for the lower Co contents. As shown in Fig.4(b) for $x\gtrsim4$\%, this yields $1/\tau(T)$ data which
split, as usual, in a $T$-dependent $1/\tau_{e}(T)$ and a
residual $1/\tau_{0}$. The former is very well
fitted below 150K by a Fermi liquid like $T^{2}$ law highly dependent on $x$, which excludes an electron-phonon scattering contribution. Such electron-electron scattering processes are given by:
\begin{equation}
\hbar /\tau_{e}(T)=A(k_{B}T)^{2}/E_{F}
\end{equation}
where $A$ is a dimensionless constant \cite{Ashcroft}. They are overwhelmed by phonon processes in usual
metals, but are enhanced here since $k_{B}T\simeq E_{F}$. For
$x=14$\%, with $\ E_{F} = 25$meV and $m_{e}/m_{0}=2$ taken from ARPES, we estimate a typical value $A\simeq4$, which corroborates the validity of our analysis. We furthermore find that the
$T^{2}$ coefficient displays a simple $1/n_{e}$ variation (inset of Fig.4(a)), which agrees with $E_{F}\varpropto n_{e}^{\alpha}$ for a
simple parabolic band, with $\alpha=1$ in the 2D case. As for the disorder induced  residual scattering rate $\tau_{0}^{-1}$, we can see in Fig.4-b, that it remains unchanged in the paramagnetic state for all $x$, so that native disorder dominates over the effect of Co substitution in the Fe planes. 

\paragraph{Parent compound and weakly doped compounds.}
For $x=0$, the drop of $n_{H}$ in fig.2 occurs in two steps: a first one at $T_{SDW}=135$K, results in a plateau which is followed then by a slower decrease down to $T=0$. Note that this two step transition is not seen in $\rho(T)$ and that neutron scattering experiments did not reveal any modification of the magnetic order below $T_{SDW}$ \cite{Wilson}. The value $|n_{H}|\simeq0.074$ at the plateau corresponds to half of the band calculation value $n_{e}=n_{h}=0.15$ \cite{Ma}. It matches as well the extrapolation value of $|n_{H}(T)|$ (triangle in fig.3). We notice in fig.2 that the remarkable parallel $T$ variations of $|n_{H}|$ for $x=0$ and 0.02 in the paramagnetic state correspond incidentally also to an abrupt loss of about 0.075 \textit{e}/Fe. This ensemble of observations leads us to suggest that the first step of the SDW transition for $x=0$ corresponds to the loss of one electronic band. This is expected from the gap opening associated with the nesting between one electron band and a hole band. The other bands appear then to undergo a gradual less perfect nesting, as $|n_{H}|$ remains as low as 1.5\% per Fe at low $T$. We can also speculate that the similar drop of $|n_{H}|$ for $x=2$\% is linked to the disappearance of one electronic band at $E_{F}$. The band structure of BaFe$_{2}$As$_{2}$ appears then very fragile as it is disturbed by a small shift of the chemical potential. This could be related to the strong hybridization of the electron bands near the \textit{M} point of the Brillouin zone as revealed recently by ARPES measurements \cite{Yi}. Further refined ARPES studies should permit to investigate these band structure modifications revealed by our results.

\paragraph{Discusion and conclusion.}

Despite the multiband nature of the electronic structure of these
pnictides we are able here to give a coherent picture of the charge
transport in Ba(Fe$_{1-x}$Co$_{x}$)$_{2}$As$_{2}$
single crystals at low $T$. The small values of $E_{F}$ in
these compounds have been shown to induce unusually large
variations of carrier content and of the Fermi liquid electron-electron
scattering rate. The $T$ linear contributions to $\rho (T)$ seen in the data cannot be compared
with the case of single band correlated
electron systems as recently proposed \cite{Gooch, Doiron}. In the present multiband
system the joint analysis of $\rho (T)$ and $n_{H}(T)$
washes out those in $1/\tau _{e}(T)$, which does not reflect any
quantum critical behaviour.

The large variations  of $n_{H}(T)$ seen at high $T$ imply an  unexpected semiconducting
like increase of $\sigma _{h}(T)$.
This could result from a variation of $n_{e}(T)$ due to a
shift of the chemical potential $\mu (T)$ induced by localization of
the hole carriers at low $T$. These variations could be related with the large high $T$ increase of the
NMR Knight shifts $K$ \cite{Ning, Grafe} interpreted as a
pseudogap, although $K$ is  not found to go through a maximum as in
cuprates \cite{RMP}. Others \cite{Fang} suggest that the
variation of $n_{H}(T)$ is dominated by a decrease of hole mobility
due to spin fluctuations. Let us note that the critical behaviour of spin
fluctuations detected in NMR above $T_{SDW}$ \cite{Ning} are not
directly apparent in the transport. However, we can notice that the low $T$
carrier content is smaller than expected from band calculations up to $x\backsimeq10$\% and that the large
high $T$ increase in $|n_{H}(T)|$ appears mainly for the magnetic samples. In the approach of \cite{Fang}, this might mean that hole localization is promoted by spin fluctuations. Nevertheless we do not understand why electron
transport would not be affected as well in such a case. This unusual
behavior is an originality of the electronic correlations in these 
compounds which requires a specific theoretical understanding.

As for SC, it seems to coexist with the SDW state in these doped compounds,
contrary to the case of the electron doped 1111 family \cite{Zhao, Luetkens}. 
However we evidence that SC only strengthens when the SDW nesting
transition has been sufficiently weakened, which agrees qualitatively with
the suggestion done for a $(s\pm )$ superconductivity with pairing of
electrons in distinct hole and electron bands \cite{Mazin2}. Further
transport on the hole doped cases should allow to reveal any asymmetry of
the phase diagram from the point of view of transport properties.

\begin{acknowledgments}
This work has been performed within the "Triangle de la Physique". We thank S. Poissonnet (SRMP/CEA) for the chemical analyses of the samples. We acknowledge fruitful discussions with I. Mazin, V. Brouet and C. P\'epin and thank N. Kirova and P. Mendels for critical reading of the manuscript.
\end{acknowledgments}

\end{document}